\documentclass[aps,prl,twocolumn,superscriptaddress]{revtex4-1}
\usepackage{graphicx}
\usepackage{upgreek}
\begin{document}

\title{Watching the birth of a charge density wave order: \\diffraction study on nanometer-and picosecond-scales}

\author{C.~Laulh\'e}
\email[Corresponding author: ]{laulhe@synchrotron-soleil.fr}
\affiliation{Synchrotron SOLEIL, L'Orme des Merisiers, Saint Aubin - BP 48, F-91192 Gif-sur-Yvette, France}
\affiliation{Universit\'e Paris-Saclay (Univ. Paris-Sud), F-91405 Orsay Cedex, France}

\author{T.~Huber}
\affiliation{Institute for Quantum Electronics, Physics Department, ETH Zurich, CH-8093 Zurich, Switzerland}

\author{G.~Lantz}
\affiliation{Laboratoire de Physique des Solides, Universit\'e Paris-Sud, CNRS, UMR 8502, F-91405 Orsay, France}
\affiliation{Institute for Quantum Electronics, Physics Department, ETH Zurich, CH-8093 Zurich, Switzerland}

\author{A.~Ferrer}
\affiliation{Swiss Light Source, Paul Scherrer Institute, CH-5232, Villigen, Switzerland}

\author{S.O.~Mariager}
\affiliation{Swiss Light Source, Paul Scherrer Institute, CH-5232, Villigen, Switzerland}

\author{S.~Gr\"ubel}
\affiliation{Swiss Light Source, Paul Scherrer Institute, CH-5232, Villigen, Switzerland}

\author{J.~Rittmann}
\affiliation{Swiss Light Source, Paul Scherrer Institute, CH-5232, Villigen, Switzerland}

\author{J.A.~Johnson}
\affiliation{Swiss Light Source, Paul Scherrer Institute, CH-5232, Villigen, Switzerland}

\author{V.~Esposito}
\affiliation{Swiss Light Source, Paul Scherrer Institute, CH-5232, Villigen, Switzerland}

\author{A.~L\"ubcke}
\altaffiliation[Present address: ]{Max-Born Institute for Nonlinear Optics and Short Pulse Spectroscopy, Max-Born-Strasse 2A, 12489 Berlin, Germany}
\affiliation{Swiss Light Source, Paul Scherrer Institute, CH-5232, Villigen, Switzerland}

\author{L.~Huber}
\affiliation{Institute for Quantum Electronics, Physics Department, ETH Zurich, CH-8093 Zurich, Switzerland}

\author{M.~Kubli}
\affiliation{Institute for Quantum Electronics, Physics Department, ETH Zurich, CH-8093 Zurich, Switzerland}

\author{M.~Savoini}
\affiliation{Institute for Quantum Electronics, Physics Department, ETH Zurich, CH-8093 Zurich, Switzerland}

\author{V.L.R.~Jacques}
\affiliation{Laboratoire de Physique des Solides, Universit\'e Paris-Sud, CNRS, UMR 8502, F-91405 Orsay, France}

\author{L.~Cario}
\affiliation{Institut des Mat\'eriaux Jean Rouxel - UMR 6502, Universit\'e de Nantes, 2 rue de la Houssini\`ere, F-44322 Nantes, France}

\author{B.~Corraze}
\affiliation{Institut des Mat\'eriaux Jean Rouxel - UMR 6502, Universit\'e de Nantes, 2 rue de la Houssini\`ere, F-44322 Nantes, France}

\author{E.~Janod}
\affiliation{Institut des Mat\'eriaux Jean Rouxel - UMR 6502, Universit\'e de Nantes, 2 rue de la Houssini\`ere, F-44322 Nantes, France}

\author{G.~Ingold}
\affiliation{Swiss Light Source, Paul Scherrer Institute, CH-5232, Villigen, Switzerland}

\author{P.~Beaud}
\affiliation{Swiss Light Source, Paul Scherrer Institute, CH-5232, Villigen, Switzerland}

\author{S.L.~Johnson}
\affiliation{Institute for Quantum Electronics, Physics Department, ETH Zurich, CH-8093 Zurich, Switzerland}

\author{S.~Ravy}
\affiliation{Laboratoire de Physique des Solides, Universit\'e Paris-Sud, CNRS, UMR 8502, F-91405 Orsay, France}

\date{\today}

\begin{abstract}
Femtosecond time-resolved X-ray diffraction is used to study a photo-induced phase transition between two charge density wave (CDW) states in 1T-TaS$_2$, namely the nearly commensurate (NC) and the incommensurate (I) CDW states. Structural modulations associated with the NC-CDW order are found to disappear within 400~fs. The photo-induced I-CDW phase then develops through a nucleation/growth process which ends 100~ps after laser excitation. We demonstrate that the newly formed I-CDW phase is fragmented into several nanometric domains that are growing through a coarsening process. The coarsening dynamics is found to follow the universal Lifshitz-Allen-Cahn growth law, which describes the ordering kinetics in systems exhibiting a non-conservative order parameter.
\end{abstract}

\maketitle

Among strongly correlated electron systems, superconductors and materials exhibiting metal-insulator transitions are usually characterized by strong electron-electron and electron-phonon couplings \cite{Dag1994,Mas1998,Gru1994}. At thermodynamic equilibrium, the corresponding many-body interactions lead to rich phase diagrams as a function of temperature, pressure or doping. Such compounds also display fascinating out-of-equilibrium physics, in the form of ultra-fast symmetry changes known as photo-induced phase transitions \cite{Bea2009,Hub2014,Cav2001}, and occurrence of new, transient states \cite{Cav2001,Bau2007,Lan2017}.

Charge density wave (CDW) states are broken symmetry states of metals arising from electron-phonon interactions. They are characterized by a periodic modulation of both atomic positions and electron density. The metal-to-CDW phase transition is characterized by the growth of a complex-valued order parameter $p=A\exp^{i\Phi}$, which reflects both the amplitude $A$ and the phase $\Phi$ of the periodic modulation \cite{Gru1994}. A number of photo-induced phase transitions that have been achieved in CDW compounds correspond to a suppression of the CDW order, i.e. a transition between a CDW state and a metallic state free of any structural modulation \cite{Sch2008,Eic2010,Roh2011,Moh2011,Hel2012,Era2012,Zhu2013,Hub2014}. Among those, the photo-induced suppression of the CDW state in blue bronze was shown to involve a coherent motion of atoms along the normal coordinates of the CDW amplitude mode \cite{Hub2014}. In this case, the amplitude mode allows \emph{continuous} variations of the modulus of the order parameter $\left|p\right|$, the metallic state corresponding to $\left|p\right|$=0. In the present work, we focus on the photo-induced phase transition between the nearly commensurate (NC) and the incommensurate (I) CDW states in 1T-TaS$_2$, which exhibit two distinct order parameters. When thermally-induced, this first-order phase transition involves a \emph{discontinuous} change of atomic positions, and a coexistence of NC and I phase domains over a 3~K range \cite{Van1980,Ish1991}. It is thus expected that the photo-induced I phase appears through non-coherent atomic motions, by a nucleation/growth process. We report that the photo-induced NC $\rightarrow$ I phase is completed within 100~ps after laser excitation. At this 100~ps delay, the photo-induced I-CDW phase is found divided into domains with a typical size of 150~{\AA}. Its ordering kinetics could be captured, in the form of a coarsening of the domain pattern. To our knowledge, this constitutes the first experimental observation of a coarsening phenomenon on the nanometer and picosecond length- and time-scales.

\begin{figure}
\includegraphics[width = 8.5 cm]{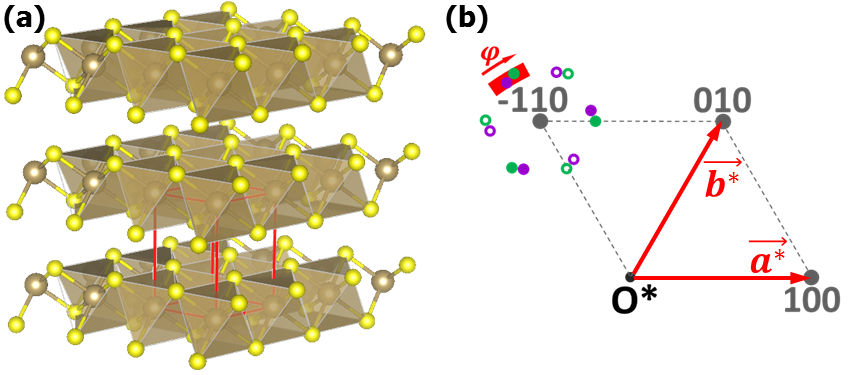}
\caption{\textbf{(a)}~Crystal structure of 1T-TaS$_2$. The hexagonal unit-cell is represented in red \cite{VESTA}. \textbf{(b)}~Location of diffracted intensity in reciprocal space. Satellite peaks related to the $\bar{1} 1 0$ lattice peak are represented in purple for the NC phase and in green for the I phase (filled circles: $l=\frac{1}{3}$, open circles: $l = -\frac{1}{3}$). In the chosen experimental conditions, scanning the azimuthal angle $\varphi$ allows measuring diffracted intensity along the thick red line.}
\label{StructDiff}
\end{figure}

1T-TaS$_2$ is a lamellar compound formed by sheets of edge-linked TaS$_6$ octahedra (Fig.~\ref{StructDiff}a). In the structure of highest symmetry, the Ta-atoms form a regular hexagonal lattice. Below 543~K, a transition to the triple-q modulated I-CDW phase occurs. In this phase, an atom which lies at the position $\vec{r}$ of the hexagonal lattice is displaced by the vector $\vec{u}\left(\vec{r}\right)=\sum_{i=1}^3{u_I}\vec{e_i}\times\cos\left(\vec{q_I}^i.\vec{r}+\Phi_I\right)$. The I-CDW modulation is characterized by its wavevector $\vec{q_I}^1=0.283\vec{a^*}+\frac{1}{3}\vec{c^*}$ and equivalents by the 3-fold symmetry \cite{Ish1991}, as well as by the order parameter $p_I={u_I}\exp^{i\Phi_I}$ \cite[p. 175]{Mot1986}.  Below 350~K, the modulation wavevectors suddenly rotate by about 12$^{\circ}$ in the $(\vec{a^*},\vec{b^*})$ plane, marking the onset of the NC-CDW state. At 300~K, the NC phase exhibits modulation vectors $\vec{q_{NC}}^1=0.245\vec{a^*}+0.068\vec{b^*}+\frac{1}{3}\vec{c^*}$ and equivalents by the 3-fold symmetry \cite{Ish1991,Spi1997}.

The advent of setups dedicated to time-resolved diffraction on sub-ps timescales has allowed detailed analyses on the mechanism of several photo-induced phase transitions \cite{Cav2001,Bau2007,Hub2014,Eic2010,Moh2011,Bea2014,Hau2016,Mar2012}. Diffraction techniques are especially well adapted to study CDW compounds \cite{Eic2010,Moh2011,Era2012,Tzo2012,Zhu2013,Hub2014,Lau2015,Sin2015,Sun2015,Hau2016,Moo2016,Sin2016}. Indeed, a structural modulation with wavevector $\vec{q}$ gives rise to satellite peaks located at positions $\pm\vec{q}$ with respect to each regular lattice peak (Fig.~\ref{StructDiff}b). Their intensity is proportional to the square of the atomic displacement amplitude, which offers a direct measurement of the CDW order parameter.

Femtosecond pump-probe diffraction experiments were carried out using a hard x-ray synchrotron slicing source \cite{Bea2007}. A platelet-like, (001)-oriented 1T-TaS$_2$ single crystal \cite{Rav2012} was excited with 1550~nm laser pulses with $p$ polarization, at an incidence angle of 10$^{\circ}$ with respect to the surface plane. The diffraction was studied in a grazing incidence geometry by using 7.05~keV, 140~fs x-ray pulses at an incidence angle of about 1$^{\circ}$. The effective penetration depths of the laser and x-ray beams \footnote{The term penetration depth here refers to the depth at which the intensity decays to 1/$e$ of its surface value.} are estimated to $\delta_L$~=~44~nm and $\delta_{RX}$~=~130~nm, respectively \cite{Bea1975,Hen1993}. In order to maintain the grazing incidence angle, the diffraction condition is tuned by rotating the sample about its surface normal. In the following, this rotation is referred to as azimuthal angle and denoted $\varphi$. The temperature was controlled between 240~K and 300~K by means of a N$_2$-blower.

\begin{figure}
\includegraphics[width = 8.5 cm]{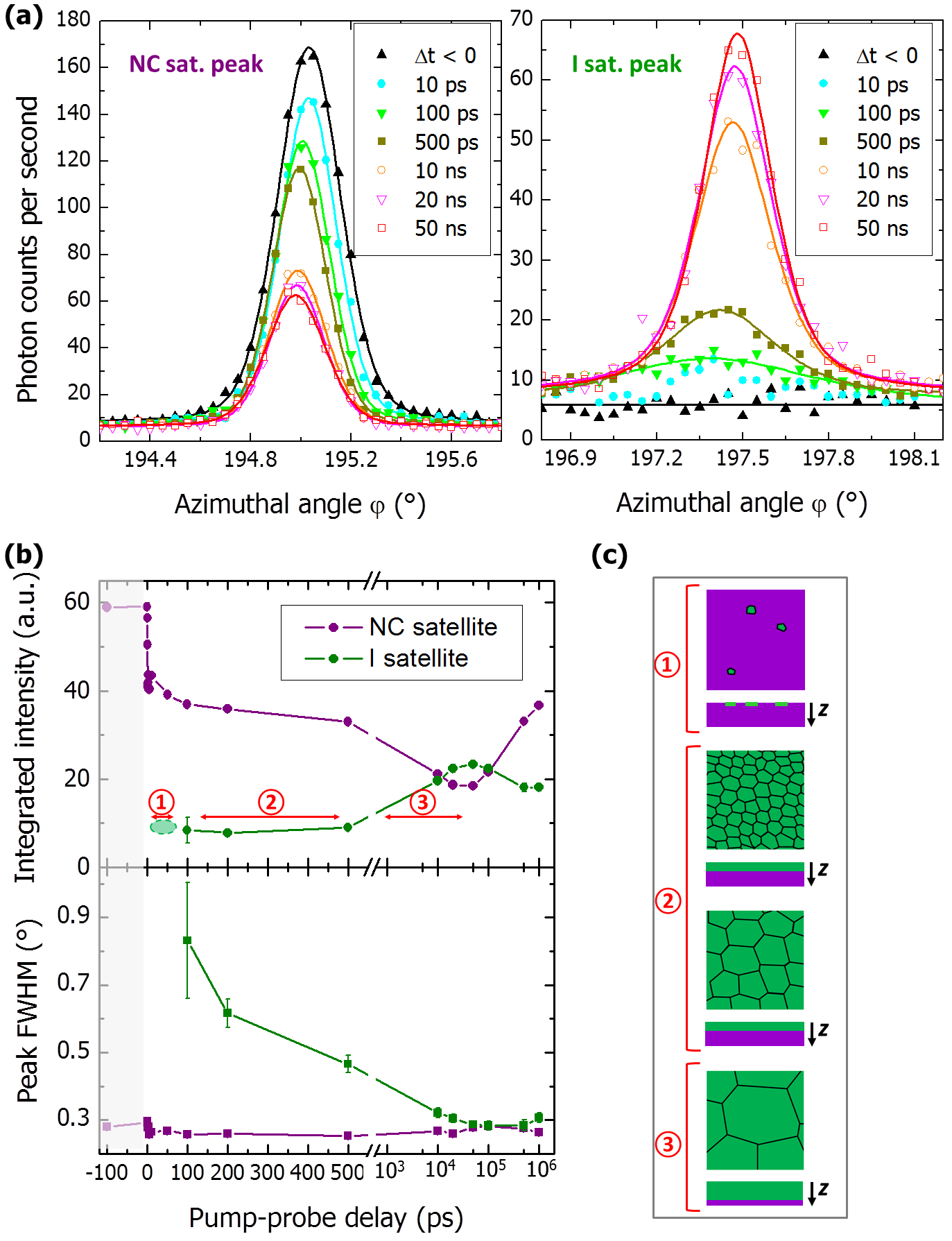}
\caption{\textbf{(a)}~Diffracted intensity profiles measured at the NC and I satellite peak positions during the NC $\rightarrow$ I photo-induced phase transition of 1T-TaS$_2$ (absorbed fluence 6.8 mJ/cm$^2$, 265~K). The dots represent measured data and the solid lines their best fit using a pseudo-Voigt function. \textbf{(b)}~Time-dependence of the NC and I peak profile parameters, as extracted from the fit. \textbf{(c)}~Schematic drawing of the 3-steps dynamics of the photo-induced NC $\rightarrow$ I phase transition. For each step, crystal views from both the top and the side are given.}
\label{PIPT}
\end{figure}

Figure~\ref{PIPT}a shows the profiles of diffracted intensity measured at the $\vec{q_{NC}}~=~(-1.313,1.245,0.333)$ and $\vec{q_{I}}~=~(-1.283,1.283,0.333)$ satellite peak positions in reciprocal space, for various time intervals after photo-excitation (absorbed fluence 6.8~mJ/cm$^2$ \footnote{The term absorbed fluence refers to the laser pulse energy per surface unit which is absorbed at the air-crystal interface. It is calculated from the optical reflectivity data published in~\cite{Bea1975}.}, 265~K). The NC~$\rightarrow$~I photo-induced transition is evidenced by the appearance of the I~satellite peak and its growth at the expense of the NC satellite peak, on timescales ranging from hundreds of fs to tens of ns. The time-dependence of the integrated intensity and full width at half maximum (FWHM) was determined for both the NC and I satellite peaks, by fitting a pseudo-Voigt function to the data (Fig.~\ref{PIPT}b). A 3-step mechanism of the photo-induced NC~$\rightarrow$~I phase transition is revealed (Fig.~\ref{PIPT}c):

1) An ultrafast NC~$\rightarrow$~I phase transition occurs in the first few~ps after laser excitation, as shown by the drop of the NC~satellite peak intensity and the concomitant appearance of diffuse scattering at the I~satellite peak position.

2) For pump-probe delays ranging from 100 to 500~ps, the integrated intensities of the NC and I~satellite peaks remain fairly constant, meaning that the modulation amplitudes and the relative volumes of the NC and I phases are stabilized. The width of the I~satellite peak, however, still evolves, giving evidence for structural rearrangements within the photo-induced I-CDW state.

\begin{figure}
\includegraphics[width = 8.5 cm]{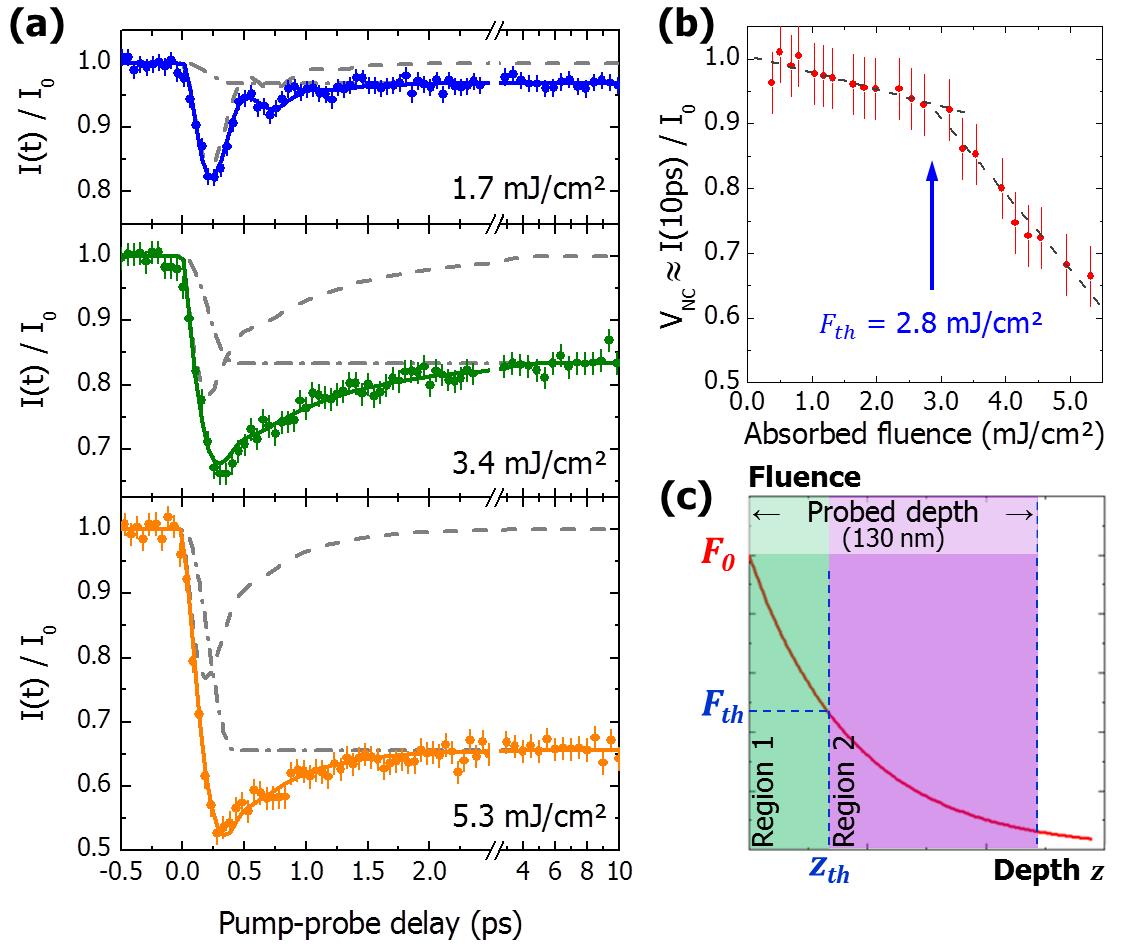}
\caption{\textbf{(a)}~Time evolution of the normalized diffracted intensity at the NC satellite peak position $-1.313~1.245~0.333$ (240~K). Lines represent the best fits using a weighted sum of the model functions defined for regions 1 and 2 (see text). \textbf{(b)}~Relative change of intensity observed for the NC satellite peak as a function of fluence ($\Delta t$~=~10~ps). \textbf{(c)}~Schematic representation of the inhomogeneous excitation across the probed depth of the 1T-TaS$_2$ crystal. The $z$ axis lies normal to the crystal's surface.}
\label{To10ps}
\end{figure}

3) A second growth of the I phase at the expense of the NC phase is observed at pump-probe delays longer than 500~ps and up to 50~ns. Taking into account the typical heat diffusivity in solids [10$^{-6}$~m$^2$.s$^{-1}$] and the 130~nm probed depth, the latter timescales can be associated with heat diffusion processes. Due to the limited penetration depth of the infrared photons, the laser excitation density decays exponentially within the probed depth of the sample (Fig. \ref{To10ps}c). The photo-induced I~phase is thus expected to nucleate close to the sample's surface. At a delay of 100 ps after laser excitation, an effective local temperature can be defined, which follows an in-depth profile similar to the one of laser excitation density. Temperature homogeneity is then slowly restored through heat diffusion, which causes a transient temperature increase in the furthest regions from the surface and, in turn, a thermally activated growth of the I phase towards sample's depth.

Figure~\ref{To10ps}a shows the time evolution of diffracted intensity at the NC satellite peak position $\vec{q_{NC}}$ (240~K, absorbed fluences ranging from 1.7 to 5.3 mJ/cm$^2$). A drop of diffracted intensity is observed, followed by a \textsl{partial} recovery within 3~ps. The diffracted intensity then remains approximately constant up to at least 10~ps. At the lowest fluence studied, periodic oscillations could be clearly resolved within 2~ps after laser excitation. Variations of the NC satellite peak intensity can reflect either a change of the NC-CDW modulation amplitude, or a change of the relative volume of the NC phase. Fig.~\ref{PIPT}a shows that 10 ps after laser excitation, the reduced intensity of the NC satellite peak is observed concomitant with a diffraction contribution located at the I satellite peak position. For larger pump-probe delays, the NC and I satellite peak intensities show an inverse correlation (Fig.~\ref{PIPT}b). These observations suggest that the intensity variations observed beyond 10~ps after laser excitation are due to changes of the relative volumes of the NC and photo-induced I phases, rather than a change of the NC or I order parameters. Under this assumption, the NC satellite peak intensity at 10~ps pump-probe delay is a measure of the relative volume of the NC phase. We write $V_{NC} \approx I\left(10~\rm{ps}\right)/I_0$, where $I_0$ denotes the diffracted intensity before laser excitation. Its fluence-dependence is reported in Fig.~\ref{To10ps}b, giving evidence for two different regimes of the photo-induced response. Below the threshold fluence $F_{th}$~=~2.8~mJ/cm$^2$, $V_{NC}$ slowly decreases with fluence, but remains greater than 93\%. The slope then dramatically increases above $F_{th}$. We propose that $F_{th}$ is the threshold fluence of the photo-induced NC~$\rightarrow$~I phase transition \cite{Moh2011}.

Assuming linear absorption, the effective laser fluence decays exponentially across the probed depth $z$, following $F(z) = F_0~e^{-z/\delta_L}$. In the case where $F_0 > F_{th}$, the sample splits into two regions exhibiting each a different behavior upon laser excitation (Fig.~\ref{To10ps}c):

\emph{Region 1-} [$z < z_{th} = \delta_L~ln(F_0 / F_{th})$] This region undergoes the photo-induced NC~$\rightarrow$~I phase transition. The NC satellite peak intensity measured from this region is expected to drop to zero within few~ps, as a result of both a decrease of the NC order parameter and a reduction of the relative volume of the NC phase. We chose to model the disappearance of the NC satellite peak intensity by a sigmoid-shaped function. One writes $\frac{I^1(t)}{I^1_{0}} = S(t)$, where $S(t)$ equals $\frac{1}{2} \left(1 + \cos\left[\frac{\pi t}{T_s}\right]\right)$ when $0 \leq t \leq T_s$, and 0 when $t > T_s$. $T_s$ is the completion time of suppression of the NC-CDW order in region 1.

\emph{Region 2-} [$z > z_{th}$] No photo-induced transition occurs in this region. In the low fluence regime, laser pulses are known to coherently excite the amplitude mode of the NC-CDW \cite{Dem2002,Per2006,Per2008}, which results in periodic oscillations of the modulus of the order parameter $p_{NC}$ in time. This behavior can be modeled assuming a displacive excitation mechanism \cite{Zei1992,Hub2014}. For $t \geq 0$, one writes $\frac{I^2(t)}{I^2_{0}} = \left(1 + A_d \left[\cos\left(2\pi\nu_{AM}t\right)e^{-t/\tau_{AM}} - e^{-t/\tau_d}\right]\right)^2$, where $A_d$, $\nu_{AM}$ and $\tau_{AM}$ represent the amplitude, frequency, and damping time of the coherent oscillations of the amplitude mode. The time constant $\tau_d$ characterizes the relaxation of the transient quasi-equilibrium atomic positions.

The function $V_{NC}\frac{I^1(t)}{I^1_{0}} + \left(1-V_{NC}\right)\frac{I^2(t)}{I^2_{0}}$ allows an excellent fit of the experimental data (Fig.~\ref{To10ps}a). In sample parts subjected to a fluence lower than $F_{th}$, the coherently excited amplitude mode is found to exhibit a frequency $\nu_{AM}$ of 1.9$\pm$0.2~THz and a short damping constant ($\tau_{AM}$ varies from 290 to 140~fs as $F_0$ increases from 1.7 to 5.3~mJ/cm$^2$). The parameter $\tau_d$ is found equal to 570$\pm$200~fs at all fluences studied. These results are consistent with those of previous time-resolved ARPES and optical measurements \cite{Dem2002,Per2006,Per2008}. On the other hand, in sample parts subjected to a fluence higher than $F_{th}$, the fit using the $S(t)$ function shows that NC modulations decrease within $T_s$~=~400~fs, slower than half a period of the NC-CDW amplitude mode (Fig. \ref{To10ps}c). A plausible scenario behind this observation could be a displacive excitation of an \emph{overdamped} amplitude mode, as already proposed in Ref.~\onlinecite{Hau2016}.

\begin{figure}
\includegraphics[width = 8.5 cm]{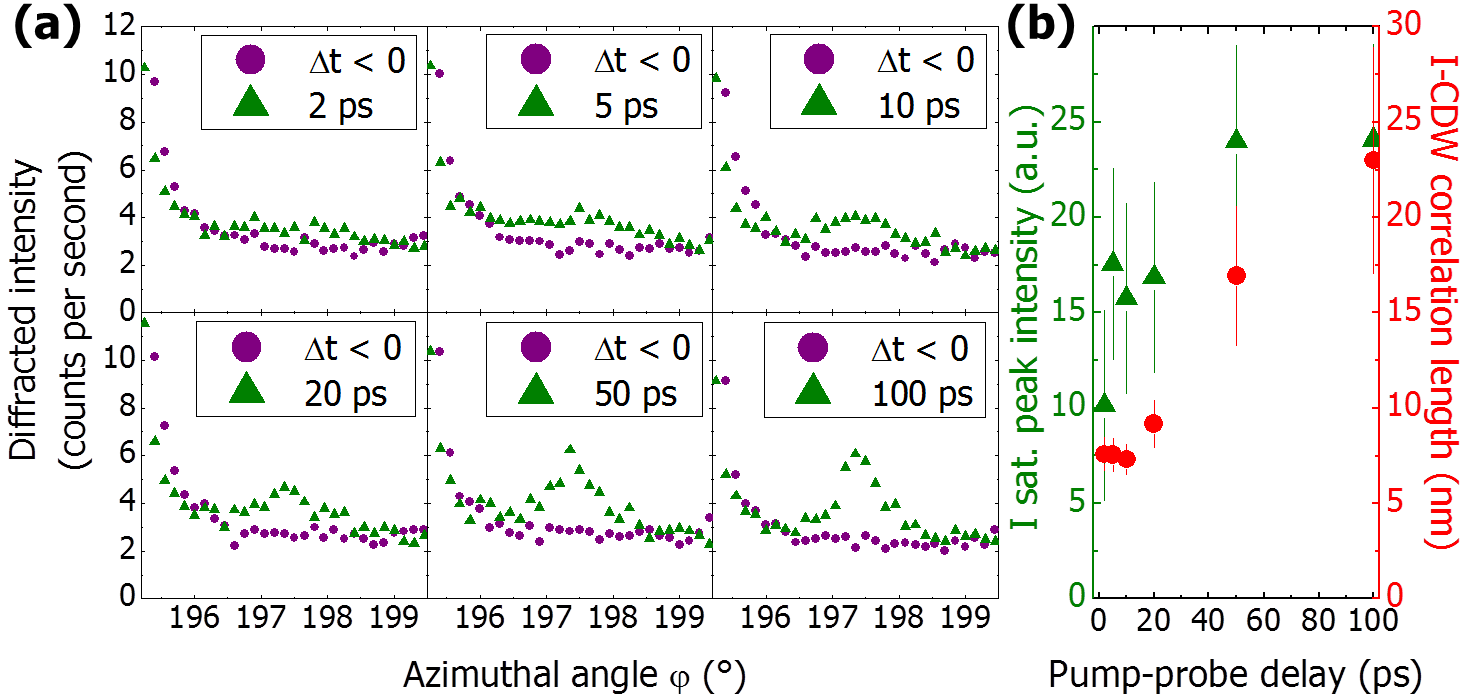}
\caption{\textbf{(a)}~Diffraction profiles of the photo-induced I satellite peak ($F$~=~6.8~mJ/cm$^2$, $T$~=~240~K). The strong contribution observed on the small angle side corresponds to the edge of the NC satellite peak. \textbf{(b)}~Time-evolutions of the integrated intensity of the I satellite peak (triangular dots), and of the photo-induced I-CDW correlation length (round dots).}
\label{PeakI}
\end{figure}

Figure~\ref{PeakI}a shows X-ray diffraction profiles of the I satellite peak, measured at various delays after laser excitation ($F_0$~=~6.8~mJ/cm$^2$, T~=~240~K). Their angular FWHMs ($\sim$1.7$^{\circ}$ and $\sim$0.6$^{\circ}$ at $\Delta t$~=~2~ps and 100~ps, respectively) are found remarkably larger than those of the NC satellite peaks ($\sim$0.3$^{\circ}$, see Fig.~\ref{PIPT}b) and regular lattice peaks ($\sim$0.3$^{\circ}$, data not shown). In the geometry used for our diffraction experiment, the broadening of the I satellite peak $\Delta\varphi_I$ \footnote{Assuming an instrumental width of 0.3$^{\circ}$, the broadening is calculated as $\Delta\varphi_I~=~\sqrt{FWHM^2 - 0.3^2}$.} is related to the spread of diffracted intensity along the $\left[\vec{a^*}+\vec{b^*}\right]$ direction (see~Fig.~\ref{StructDiff}b): $\Delta q_{I} \approx \left\|\vec{q_{I}}-\frac{1}{3}\vec{c^*}\right\|\Delta\varphi_I$. $\Delta q_{I}$ finally translates into the correlation length of the photo-induced I-CDW in the $(\vec{a},\vec{b})$ plane, following $\xi_{I}={2\pi}/{\Delta q_{I}}$. It is found to increase from 7 to 23~nm in the [2-100~ps] delay range (Fig.~\ref{PeakI}b). In parallel, we estimated the integrated intensity of the photo-induced I satellite peaks, by summing the positive count differences $I\left(t\right)-I\left(t<0\right)$ over the $\varphi$-angle range (Fig.~\ref{PeakI}b). The diffracted intensity is found to progressively increase in the pump-probe delay range [2-100~ps], indicating an increase of the photo-induced I phase volume. The concomitant growth of correlation length and increase of the I phase volume observed is fully compatible with the expected scenario of a first order phase transition, where nuclei of the new phase are produced and grow over time. Note that the study of the weak I satellite peak at short delays was possible only at 240~K: at higher temperatures, the background of thermal diffuse scattering dominates the diffracted signal (Fig.~\ref{PIPT}a).

\begin{figure}
\includegraphics[width = 8.5 cm]{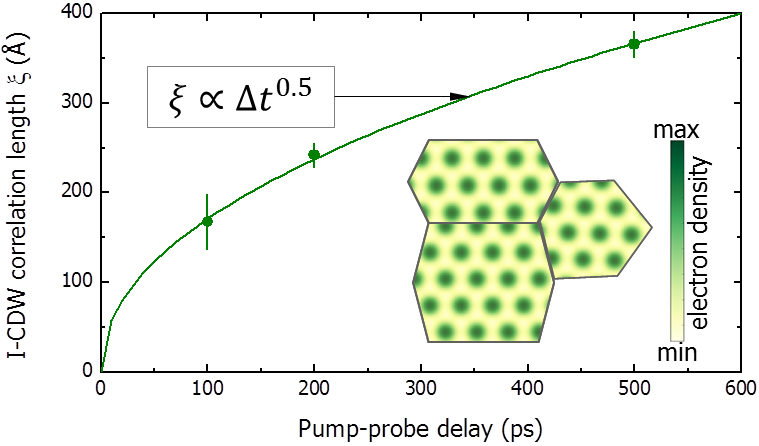}
\caption{\textbf{(dots)}~Time evolution of the correlation length of the photo-induced I-CDW phase ($F$~=~6.9~mJ/cm$^2$, $T$~=~265~K). \textbf{(line)}~Best fit using the power law function $C \times \Delta t^{0.5}$. The bottom right insert represents a subset of 3 I-CDW domains, each characterized by a different CDW phase $\Phi_{In}$. Modulations of electronic density associated with the I-CDW are color-coded within each of the 3 domains. For the sake of simplicity, sharp domain walls are represented: their actual width cannot be deduced from the present experiment.}
\label{Correl}
\end{figure}

Let us now have a closer look on the experimental data presented in Fig.~\ref{PIPT}b ($F_0$~=~6.9mJ/cm$^2$, T~=~265~K). In the pump-probe delay range [100-500~ps], the progressive reduction of the I satellite peak FWHM is still observed, meaning that the correlation length of the I-CDW is still increasing. On the other hand, the integrated intensity of the I satellite peak remains constant, meaning that the volume of the photo-induced I phase is stabilized. These observations exclude a scenario where the growth of correlation length would be due to the growth of I phase regions. Instead, the growth of correlation length should be thought as an internal rearrangement \emph{within} the photo-induced I-CDW phase. The correlation length of the I-CDW state is a measure of the typical distance over which the phase of the modulation remains constant. We thus assume that the photo-induced I phase is fragmented into small domains, each exhibiting one phase $\Phi_{In}$ (see inset in Fig.~\ref{Correl}).

In the time interval [100-500~ps] after laser excitation, both the electron and phonon energy distributions are expected to be thermalized with a single characteristic local temperature $T\left(z\right)$. The observed development of the I phase beyond 100~ps pump-probe delay gives evidence that $T\left(z\right)$ is greater than the critical temperature of the NC $\rightarrow$ I phase transition, at least for some depths $z$ close to sample's surface \footnote{$T\left(z\right)$ is expected to exhibit a similar $z$-dependence as $F\left(z\right)$ (see Fig.~\ref{To10ps}c).}. As a consequence, the free energy is well minimized within each photo-induced I-CDW domain. The phases $\Phi_{In}$ can in principle take any value, owing to the incommensurability of the I-CDW with the underlying hexagonal lattice. Nevertheless, the steep change in the I-CDW phase which occurs through the domain walls increases the energy of the system. This leads to a coarsening dynamics of the domain pattern, driven by a reduction of the domain wall area (Fig.~\ref{PIPT}c, thumbnail 2).

Coarsening systems are known to obey the dynamic scaling hypothesis \cite{Bra2002}. Under such an assumption, the domain pattern at later times exhibits similar length distributions as at earlier times, provided that a global scale factor is applied. As a consequence, the time-dependent growth of the domains has to be modeled by a scale-invariant power law $At^p$. Figure~\ref{Correl} shows the time evolution of the I-CDW correlation length in the [100-500~ps] delay range, as deduced from the broadening of the I satellite peak. The time-dependent power law allows an excellent fit of the data, the refined value for the exponent $p$ being 0.47$\pm$0.03. The coarsening dynamics of the photo-induced I-CDW is found to follow the universal Lifshitz-Allen-Cahn growth law ($t^{1/2}$), which describes domain growth in systems where the order parameter is not conserved \cite{Bra2002}. Note that this behavior has seldom been observed experimentally, exclusively until now on quenched liquid crystals, on micrometer and second length- and time-scales~\cite{Ori1986,Sic2008}.

In summary, we depicted the photo-induced structural dynamics in the NC phase of 1T-TaS$_2$, over the relevant range of timescales [100~fs~-~1~$\upmu$s]. At moderate excitation fluences, we provided the first direct experimental evidence of a displacive excitation of the NC-CDW amplitude mode. The photo-induced NC $\rightarrow$ I phase transition occurs for absorbed laser fluences above 2.8~mJ/cm$^2$ at 240~K. In this high excitation regime, the NC-CDW modulations completely disappear within 400~fs. Regions exhibiting I-CDW modulations nucleate and grow within the first 100 ps after laser excitation. At longer pump-probe delays ($\geq$ 100~ps), the photo-induced I phase has fully developed close to the sample's surface. Nonetheless, it does not correspond to the I phase observed at thermodynamic equilibrium: its short correlation length ($\sim$15~nm) makes it a genuine out-of-equilibrium state. Some of the photo-induced nanometric domains grow at the expense of others, in a coarsening process driven by a reduction of the domain wall area. This peculiar process explains the unusually late observation of an out-of-equilibrium electronic state, in a system where electron and phonon distributions already reached a local equilibrium. The present work calls for further studies on the fast domain wall dynamics in broken-symmetry phases, which has important implications on the mechanisms of (photo-induced) phase transitions, as well as on material responses to an external field or stimulus.

\begin{acknowledgments}
We wish to warmly thank Sabrina Salmon for her valuable help during sample synthesis, as well as Daniel Grolimund for sharing his expertise during beamline alignment operations. The time-resolved x-ray diffraction measurements were performed on the X05LA beam line at the Swiss Light Source, Paul Scherrer Institut, Villigen, Switzerland. Preparative static grazing incidence diffraction measurements were performed at the CRISTAL beamline of SOLEIL synchrotron, Saint-Aubin, France. The research leading to these results has received funding from the European Community's Seventh Framework Programme (FP7/2007-2013) under grant agreement 312284 (CALIPSO).
\end{acknowledgments}

\bibliography{Laulhe_bibfile}

\end{document}